# Numerical simulation of time delay interferometry for eLISA/NGO

**Gang Wang[1] and Wei-Tou Ni[2,3]**

[1]Shenzhen National Climate Observatory, No.1 Qixiang Rd., Zhuzilin, Futian District, Shenzhen, 518040, China
[2]Center for Gravitation and Cosmology (CGC), Department of Physics, National Tsing Hua University, No. 101, Kuang Fu II Rd., Hsinchu, Taiwan, 300, ROC
[3]Shanghai United Center for Astrophysics (SUCA), Shanghai Normal University, 100 Guilin Road, Shanghai, 200234, China

E-mail: gwanggw@gmail.com, weitou@gmail.com

**Abstract**. eLISA/NGO is a new gravitational wave detection proposal with arm length of $10^6$ km and one interferometer down-scaled from LISA. Just like LISA and ASTROD-GW, in order to attain the requisite sensitivity for eLISA/NGO, laser frequency noise must be suppressed to below the secondary noises such as the optical path noise, acceleration noise etc. In previous papers, we have performed the numerical simulation of the time delay interferometry (TDI) for LISA and ASTROD-GW with one arm dysfunctional by using the CGC 2.7 ephemeris. The results are well below their respective limits which the laser frequency noise is required to be suppressed. In this paper, we follow the same procedure to simulate the time delay interferometry numerically. To do this, we work out a set of 1000-day optimized mission orbits of the eLISA/NGO spacecraft starting at January 1st, 2021 using the CGC 2.7 ephemeris framework. We then use the numerical method to calculate the residual optical path differences in the second-generation TDI solutions as in our previous papers. The maximum path length difference, for all configurations calculated, is below 13 mm (43 ps). It is well below the limit which the laser frequency noise is required to be suppressed for eLISA/NGO. We compare and discuss the resulting differences due to the different arm lengths for various mission proposals -- eLISA/NGO, an NGO-LISA-type mission with a nominal arm length of $2 \times 10^6$ km, LISA and ASTROD-GW.

## 1. Introduction and summary

Gravitational wave (GW) antennas are already on air for detecting GWs in the frequency band 10 Hz – 10 kHz for the ground-based interferometers and in the very low frequency band (300 pHz–100 nHz) for Pulsar Timing Arrays [PTAs] (see, e.g., Ni, 2010; Arun *et al.*, 2012). The second generation ground-based interferometers—Advanced LIGO (The Advanced LIGO Team, 2010), Advanced Virgo (The Advanced Virgo Team, 2010) and KAGRA (LCGT) (Kuroda et al., 2010; http://www.icrr.u-tokyo.ac.jp/2012/01/28161746.html)—under construction will reach sensitivities that promise a good chance for detecting GWs from binary neutron-star mergers around 2016. PTAs will seek for the detection of GW background and single events from supermassive black hole merger in the very low frequency band around 2020 (Demorest et al., 2009). CMB experiments [Planck Surveyor (2012), ACTPol (Niemack *et al*, 2010; McMahon *et al*, 2012), and SPTPol (Austermann *et al*, 2012; George *et al*, 2012), etc.] are currently online/under upgrading to search for GWs in the extremely low (Hubble) frequency band (1 aHz–10 fHz); significant progress in sensitivity will be made in 5 years with the hope of detecting GW.

In between the high frequency band and the very low frequency band, there are the middle frequency band (0.1 Hz–10 Hz) and the low frequency band (100 nHz–0.1 Hz). *Space laser-interferometric GW detectors* operate in these bands. Mission concepts under implementation/study are eLISA/NGO (http://eLISA-ngo.org/; Jennrich *et al*, 2011), Super-ASTROD (Ni, 2009a), ASTROD-GW (Ni, 2009b; Ni *et al*, 2009), BBO (Crowder and Cornish, 2005; http://universe.nasa.gov/new/program/bbo.html) and DECIGO (Kawamura *et al*, 2006, 2011; Ando and the DECIGO Working Group, 2013). The GW sources for these missions are well documented and the signal-to-noise ratios are high (LISA Study Team, 2000; Ni, 2009a; Ni, 2010, 2012, 2013; Phinney *et al*, 2004; Kawamura *et al*, 2011).

Except DECIGO whose configuration is basically like the ground GW interferometric detectors, all other laser-interferometric antennas for space detection of GWs have their arm lengths vary with time according to orbit dynamics. In order to attain the requisite sensitivity, laser frequency noise must be suppressed to below the secondary noises such as the optical path noise, acceleration noise etc. The TDI technique can be used to suppress the laser frequency noise. The basic principle of TDI is to use two different optical paths but whose optical path lengths are nearly equal, and follow them in opposite order. This operation suppresses the laser frequency noise if the two paths compared are close enough in optical path length (time travelled).

In previous papers, we have performed the numerical simulation of the time delay interferometry for LISA (Dhurandhar, Ni and Wang, 2013) and ASTROD-GW (Wang and Ni, 2013) with one arm dysfunctional by using the CGC 2.7 ephemeris. The results are well below their respective limits which the laser frequency noise is required to be suppressed. In this paper, we follow the same procedure to simulate the time delay interferometry numerically for eLISA/NGO with nominal arm length $1 \times 10^6$ km, and an NGO-LISA-type mission with a nominal arm length of $2 \times 10^6$ km. First, we briefly review these space mission concepts, the techniques of TDI for two-arm interferometer, and CGC 2.7 ephemeris.

*1.1. LISA*

LISA—Laser Interferometric Space Antenna—was a proposed ESA-NASA mission which would use coherent laser beams exchanged between three identical spacecraft (S/C) forming a nearly equilateral triangle of side $5 \times 10^6$ km inclined by about 60° with respect to the ecliptic to observe and detect low-frequency cosmic GW (LISA Study Team, 2000). The three S/C were designed to be drag-free and to trail the Earth by about 20° in an orbit around the Sun with periods about one year. This project nominally ended with NASA's withdrawal in April, 2011.

*1.2. eLISA/NGO*

eLISA/NGO is a joint effort of seven European countries (France, Germany, Italy, The Netherlands, Spain, Switzerland, UK) and ESA. The NGO assessment study report received excellent scientific evaluation (http://eLISA-ngo.org). The science objectives of eLISA/NGO are, through the detection and observation of gravitational waves, (i) to survey compact stellar-mass binaries and study the structure of the galaxy; (ii) to trace the formation, growth, and merger history of massive black holes; (iii) to explore stellar populations and dynamics in galactic nuclei; (iv) to confront General Relativity with observations and (v) to probe new physics and cosmology. The mission configuration consists of a "mother" S/C at one vertex and two "daughter" S/C at two other vertices with the mother S/C optically linked with two daughter S/C forming an interferometer. The duration of the mission is 2 years for science orbit and about 4 years including transferring and commissioning. The mission S/C orbit configuration is similar to LISA, but with nominal arm length of $1 \times 10^6$ km, inclined by about 60° with respect to the ecliptic, and trailing Earth by 10-20°.

*1.3. An NGO-LISA-type mission with a nominal arm length of $2 \times 10^6$ km*

With more funding and/or more countries participating, it would be desirable for eLISA/NGO to extend its arms for gaining sensitivities and reaching more sources. In this paper, we consider an extension of arm length to $2 \times 10^6$ km for comparison.

## 1.4. ASTROD-GW

ASTROD-GW is a dedicated mission concept for GW detection stemmed from the general concept of ASTROD (Astrodynamical Space Test of Relativity using Optical Devices) (Ni, 2008). The general of ASTROD is to have a constellation of drag-free spacecraft navigate through the solar system to range with one another using optical devices for mapping the solar-system gravitational field, for measuring the related solar-system parameters, for testing the relativistic gravity, for observing solar g-mode oscillations, and for detecting GWs. ASTROD-GW proposal has been submitted to CNSA for consideration (Ni, 2009b). The mission orbits of the 3 spacecraft forming a nearly equilateral triangular array are chosen to be near the Sun-Earth Lagrange points L3, L4 and L5. The nominal arm length is about $2.6 \times 10^8$ km (260 times that of eLISA/NGO or 52 times that of LISA).

## 1.5. *The technique of TDI for two-arm interferometer*

eLISA/NGO is a one-interferometer space mission with three S/C and two arms. In the general orbit model of eLISA/NGO, since the arm lengths vary with time, the second-generation TDIs are needed for suppressing the laser frequency noise (see, e.g., Tinto and Dhurandhar, 2005; and references therein). In this paper, we work out the time delay interferometry *numerically* for eLISA/NGO using the TDI observables listed in Dhurandhar *et al* (2010) As these TDI observables were originally worked out for LISA on the condition that only the data streams from two arms are considered, these configurations could be readily applied to the case of eLISA/NGO or similar situation. The solutions however are approximate in the sense that the higher order terms involving $\dot{L}^2$ or $\ddot{L}$ are ignored in the calculation, where $L(t)$ is the generic arm length of the space interferometer considered and the `dot' denotes derivative with respect to time.

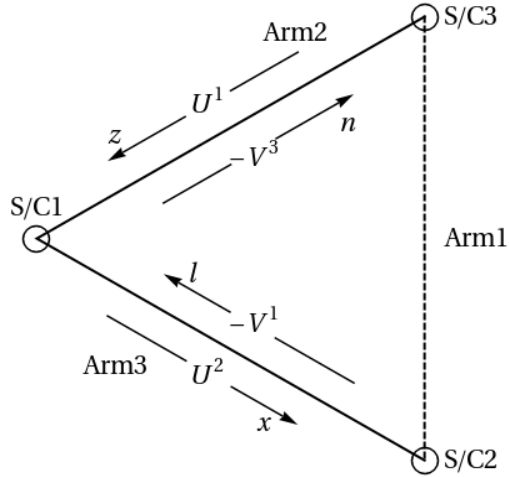

**Figure 1.** The beams and the corresponding time delays are shown schematically in the figure. The functional arms are depicted with a continuous line.

In the following, we describe the second-generation TDIs obtained by Dhurandhar *et al* (2010). Figure 1 gives a schematic description of two-arm interferometer where the four links are labelled by $U^1$, $U^2$, $V^1$, $V^3$, and the optical time-delays by $x$, $z$ (anti-clockwise) and $l$, $n$ (clockwise). Let $a = xl$ and $b = nz$ be the `round-trip' composite operators starting from S/C1, the solutions obtained by Dhurandhar *et al* (2010) are listed in degree-lexicographic order as following:

(i) n=1, $[ab, ba] = ab^2a - ba^2b$ ;
(ii) n=2, $[a^2b^2, b^2a^2]$, $[abab, baba]$, $[ab^2a, ba^2b]$;
(iii) n=3, $[a^3b^3, b^3a^3]$, $[a^2bab^2, b^2aba^2]$, $[a^2b^2ab, b^2a^2ba]$, $[a^2b^3a, b^2a^3b]$, $[aba^2b^2, bab^2a^2]$,
   $[ababab, bababa]$, $[abab^2a, baba^2b]$, $[ab^2a^2b, ba^2b^2a]$, $[ab^2aba, ba^2bab]$, $[ab^3a^2, ba^3b^2]$.

n = 3 is only an arbitrary upper limit – if desired or needed, one could go to higher degrees as well. Here and in previous papers on numerical TDIs (Dhurandhar *et al*, 2013; Wang and Ni, 2013), we use the convention that the paths start from left to right, e.g., the path *ab* starts with *a* then *b*. We note that the first-generation TDI configuration [*a*, *b*] was first considered and numerically calculated for ASTROD (Ni *et al*, 1997).

The importance of the numerical computations is that the results are obtained for more realistic spacecraft orbits that take into account the gravitational effects of most objects in the solar system including several hundred asteroids. This goes beyond semi-analytic modelling. For optical paths going up to n = 3, successive time-delays up to 23 are considered, i.e., polynomials up to degree 23 in the elementary time-delay operators are implemented. There are 14 such TDI observables, which may be deemed sufficient to carry out astrophysical observations. The results in section 4 and section 5 show that the laser frequency noises are cancelled well within the respective limits imposed by the secondary noises for eLISA/NGO and for the NGO-LISA-type mission considered.

A related important aspect is the GW response of such TDI observables. The GW response to a TDI observable may be calculated in the simplest way by assuming equal arms (the possible differences in lengths would be sensitive to frequencies outside the detector proper bandwidth). A comprehensive and generic treatment of the responses to the second-generation TDI observables can be found in Krolak *et al* (2004). As remarked, the sensitivity of the second-generation TDI observables remains essentially the same as the first-generation ones. The small differences in lengths are important for cancellation/non-cancellation of laser frequency noise and clock noise, but not for the GW response. In our case, the GW response of all these TDI observables is essentially that of the Michelson.

*1.6. CGC ephemeris*

In 1998, we started orbit simulation and parameter determination for ASTROD (Chiou and Ni, 2000a, 2000b), and worked out a post-Newtonian ephemeris of the Sun, the major planets and 3 biggest asteroids including the solar quadrupole moment. This working ephemeris was termed as CGC 1 (CGC: Center for Gravitation and Cosmology). For an improved ephemeris framework, we considered all known 492 asteroids with diameter greater than 65 km to obtain the CGC 2 ephemeris, and calculated the perturbations due to these 492 asteroids on the ASTROD spacecraft (Tang and Ni, 2000, 2002). In building the CGC ephemeris framework, we use the post-Newtonian barycentric metric and equations of motion as derived in Brumberg (1991) for solar system bodies with PPN (Parametrized Post-Newtonian) parameters $\beta$, $\gamma$. In solving a problem, one may use any coordinate system. However, in our ephemeris, we just use the equations in Brumberg (1991) with gauge parameters $\alpha = \nu = 0$ that corresponds to the harmonic gauge adopted by the 2000 IAU resolution (Soffel, 2003).

In our first optimization of ASTROD-GW orbits (Men *et al*, 2009, 2010), we used the CGC 2.5 ephemeris in which only 3 biggest minor planets are taken into account, but the Earth's precession and nutation are added; the solar quadratic zonal harmonic and the Earth's quadratic to quartic zonal harmonic are also included.

In our recent orbit simulation of the ASTROD I proposed to ESA (Braxmaier *et al*, 2010) and in our studies of TDIs for LISA (Dhurandhar, Ni and Wang, 2013) and for ASTROD-GW (Wang and Ni, 2011, 2012, 2013; Wang, 2011), we added the perturbation of additional 349 asteroids to the CGC 2.5 ephemeris and called it the CGC 2.7 ephemeris. (The difference between the CGC 2.7 ephemeris and the CGC 2 ephemeris is that we have 352 asteroids instead of 492 asteroids, but the Earth's precession and nutation are added and the solar quadratic zonal harmonic and the Earth's quadratic to quartic zonal harmonic are also included.) For more discussions on the CGC 2.7 ephemeris, please see Wang and Ni (2011, 2012).

In this paper, we use the CGC 2.7 ephemeris to obtain optimized NGO orbits and numerically evaluate NGO TDIs. The differences in orbit evolution of Earth calculated using CGC 2.7 compared with that of DE405 starting at January 1st, 2021 for 3700 days are less than 135 m, 1.9 mas and 0.65 mas for radial distance, longitude and latitude respectively.

*1.7. Comparison of TDIs for interferometers with different arm lengths*

In table 1, we compile and compare the resulting differences for the TDIs listed in the subsection *1.5*. {(i), (ii) and (iii)} due to different arm lengths for various mission proposals -- eLISA/NGO, an NGO-LISA-type mission with a nominal arm length of $2 \times 10^6$ km, LISA and ASTROD-GW.

**Table 1.** Comparison the resulting differences due to arm lengths for various mission proposals -- eLISA/NGO, an NGO-LISA-type mission with a nominal arm length of $2 \times 10^6$ km, LISA and ASTROD-GW.

| TDI configuration | | TDI path difference $\Delta L$ | | | |
|---|---|---|---|---|---|
| | | eLISA/NGO [ps] (This work) | NGO-LISA-type With $2 \times 10^6$ km arm length [ps] (This work) | LISA [ps] (Dhurandhar, Wang and Ni, 2013) | ASTROD-GW [ns] (Wang and Ni, 2013) |
| Duration | | 1000 days | 1000 days | 1000 days | 20 years |
| n=1 | [$ab$, $ba$] | -1.5 to +1.5 | -11 to +12 | -70 to +80 | -50 to +50 |
| n=2 | [$a^2b^2$, $b^2a^2$] | -11 to +12 | -90 to +100 | -600 to +650 | -400 to +400 |
| | [$abab$, $baba$] | -6 to +6 | -45 to +50 | -300 to +340 | -200 to +200 |
| | [$ab^2a$, $ba^2b$] | -0.0032 to +0.0034 | -0.0036 to +0.004 | -0.015 to +0.013 | -0.14 to +0.11 |
| n=3 | [$a^3b^3$, $b^3a^3$] | -40 to +42 | -300 to +320 | -2,000 to +2,200 | -1,300 to +1,300 |
| | [$a^2bab^2$, $b^2aba^2$] | -30 to +32 | -220 to +260 | -1,500 to +1,800 | -1,100 to +1,100 |
| | [$a^2b^2ab$, $b^2a^2ba$] | -22 to +24 | -160 to +180 | -1,000 to +1,300 | -750 to +750 |
| | [$a^2b^3a$, $b^2a^3b$] | -13 to +14 | -100 to +110 | -600 to +750 | -450 to +450 |
| | [$aba^2b^2$, $bab^2a^2$] | -22 to +24 | -160 to +180 | -1,000 to +1,300 | -750 to +750 |
| | [$ababab$, $bababa$] | -13 to +14 | -100 to +110 | -600 to +750 | -450 to +450 |
| | [$abab^2a$, $baba^2b$] | -4.5 to +4.8 | -32 to +38 | -200 to +250 | -150 to +150 |
| | [$ab^2a^2b$, $ba^2b^2a$] | -4.5 to +4.8 | -32 to +38 | -200 to +250 | -150 to +150 |
| | [$ab^2aba$, $ba^2bab$] | -4.8 to +4.5 | -38 to +32 | -250 to +200s | -150 to +150 |
| | [$ab^3a^2$, $ba^3b^2$] | -15 to +13 | -110 to +100 | -750 to +600 | -450 to +450 |
| Nominal arm length | | 1 Gm (1 Mkm) | 2 Gm | 5 Gm | 260 Gm |
| Requirement on $\Delta L$ | | 10 m (30,000 ps) | 20 m (60,000 ps) | 50 m (150,000 ps) | 500 m (1,500 ns) |

In section 2, we work out a set of 1000-day optimized mission orbits of eLISA/NGO spacecraft starting at January 1st, 2021 using the CGC 2.7 ephemeris framework. In section 3, we work out a set of 1000-day optimized mission orbits of spacecraft for an NGO-LISA-type mission starting at January 1st, 2021 using the CGC 2.7 ephemeris framework. In section 4, we obtain the numerical results pertaining to the second-generation TDIs listed in table 1 for eLISA/NGO. In section 5, we obtain the numerical results pertaining to the second-generation TDIs listed in table 1 for an NGO-LISA-type mission. In section 6, we compare and discuss the resulting differences due to different arm lengths for various mission proposals -- eLISA/NGO, an NGO-LISA-type mission, LISA and ASTROD-GW, and conclude this paper with discussion and outlook.

## 2. eLISA/NGO mission orbit optimization

The mission orbit configuration of eLISA/NGO is similar to that of LISA but with a shorter arm length and a closer distance to Earth. The distance of any two of three spacecraft must be maintained as close as possible during geodetic flight. LISA orbit configuration has been studied analytically and numerically in various previous works (Vincent and Bender, 1987; Folkner *et al*, 1997; Cutler, 1998; Hughes, 2002; Hechler and Folkner, 2003; Dhurandhar *et al*, 2005; Yi *et al*, 2008; Li *et al*, 2008). In the mission orbit optimization for eLISA/NGO, we follow the analytical procedure of Dhurandhar *et al* (2005) to make our initial choice of initial conditions and then use the CGC ephemeris to numerically optimize the orbit configuration as we have done in ASTROD-GW orbit design (Men *et al*, 2009, 2010; Wang and Ni, 2011, 2012, 2013).

*2.1. The initial choice of the eLISA/NGO initial conditions*

Let $\alpha$ be the ratio of the planned arm length $l$ of the orbit configuration to twice radius $R$ (1 AU) of the mean Earth orbit, i.e., $\alpha = l/(2R)$. There are various ways to choose the orbits of the three spacecraft so that the orbit configuration satisfying the equal arm length requirement to first order in $\alpha$. In the following, we use the ones given in Dhurandhar *et al* (2005).

Choosing the initial time $t_0$ to be JD2459215.5 (2021-Jan-1st 00:00:00), we work in the Heliocentric Coordinate System ($X$, $Y$, $Z$). $X$-axis is in the direction of vernal equinox. First, as in Dhurandhar *et al* (2005), a set of elliptical S/C orbits is defined as

$$\begin{aligned} X_f &= R(\cos\psi_f + e)\cos\varepsilon, \\ Y_f &= R(1-e^2)^{1/2}\sin\psi_f, \\ Z_f &= R(\cos\psi_f + e)\sin\varepsilon. \end{aligned} \quad (2.1)$$

where $\psi_E$ is defined to be the position angle of Earth w.r.t. the $X$-axis at $t_0$; $\varphi_0 \equiv \psi_E - 10°$; $R = 1$ AU; $e = 0.001925$; $\varepsilon = 0.00333$. The eccentric anomaly $\psi_f$ is related to the mean anomaly $\Omega(t-t_0)$ by

$$\psi_f + e\sin\psi_f = \Omega(t-t_0). \quad (2.2)$$

where $\Omega$ is defined as $2\pi/$(one sidereal year). The eccentric anomaly $\psi_f$ can be solved by numerical iteration. Define $\psi_k$ to be implicitly given by

$$\psi_k + e\sin\psi_k = \Omega(t-t_0) - 120°(k-1), \quad \text{for } k=1,2,3. \quad (2.3)$$

Define $X_{fk}$, $Y_{fk}$, $Z_{fk}$, ($k = 1.2, 3$) to be

$$\begin{aligned} X_{fk} &= R(\cos\psi_k + e)\cos\varepsilon, \\ Y_{fk} &= R(1-e^2)^{1/2}\sin\psi_k, \\ Z_{fk} &= R(\cos\psi_k + e)\sin\varepsilon. \end{aligned} \quad (2.4)$$

Define $X_{f(k)}$, $Y_{f(k)}$, $Z_{f(k)}$, ($k = 1.2, 3$), i.e., $X_{f(1)}$, $Y_{f(1)}$, $Z_{f(1)}$; $X_{f(2)}$, $Y_{f(2)}$, $Z_{f(2)}$; $X_{f(3)}$, $Y_{f(3)}$, $Z_{f(3)}$ to be

$$\begin{aligned} X_{f(k)} &= X_{fk}\cos[120°(k-1)+\varphi_0] - Y_{fk}\sin[120°(k-1)+\varphi_0], \\ Y_{f(k)} &= X_{fk}\sin[120°(k-1)+\varphi_0] + Y_{fk}\cos[120°(k-1)+\varphi_0], \\ Z_{f(k)} &= Z_{fk}. \end{aligned} \quad (2.5)$$

The three S/C orbits are (for one-body central problem) are

$$\begin{aligned} \mathbf{R}_{S/C1} &= (X_{f(1)}, Y_{f(1)}, Z_{f(1)}), \\ \mathbf{R}_{S/C2} &= (X_{f(2)}, Y_{f(2)}, Z_{f(2)}), \\ \mathbf{R}_{S/C3} &= (X_{f(3)}, Y_{f(3)}, Z_{f(3)}). \end{aligned} \quad (2.6)$$

The initial positions can be obtained by choosing $t=t_0$ and initial velocities by calculating the derivatives w.r.t. time at $t=t_0$. With the choice of $t_0 =$ JD2459215.5 (2021-Jan-1st 00:00:00), the initial conditions (states) of three spacecraft of NGO in J2000.0 solar-system-barycentric Earth mean equator and equinox coordinates are tabulated in the third column of table 2.

**Table 2.** Initial states (conditions) of 3 S/C of eLISA/NGO at epoch JD2459215.5 for our initial choice (third column), after 1st stage optimization (fourth column) and after all optimizations (fifth column) in J2000 equatorial (Earth mean equator and equinox coordinates) solar-system-barycentric coordinate system

|  |  | Initial choice of S/C initial states | Initial states of S/C after 1st stage optimization | Initial states of S/C after final optimization |
|---|---|---|---|---|
| S/C1 Position (AU) | X | -1.53222193865×10$^{-2}$ | -1.53222193865×10$^{-2}$ | -1.53221933735×10$^{-2}$ |
|  | Y | 9.23347976632×10$^{-1}$ | 9.23347976632×10$^{-1}$ | 9.23345222988×10$^{-1}$ |
|  | Z | 4.04072005496×10$^{-1}$ | 4.04072005496×10$^{-1}$ | 4.04070800735×10$^{-1}$ |
| S/C1 Velocity (AU/day) | $V_x$ | -1.71752389145×10$^{-2}$ | -1.71926502995×10$^{-2}$ | -1.71928071373×10$^{-2}$ |
|  | $V_y$ | -1.41699055355×10$^{-4}$ | -1.41837311087×10$^{-4}$ | -1.41838556464×10$^{-4}$ |
|  | $V_z$ | -6.11987395198×10$^{-5}$ | -6.12586807155×10$^{-5}$ | -6.12592206525×10$^{-5}$ |
| S/C2 Position (AU) | X | -1.86344993528×10$^{-2}$ | -1.86344993528×10$^{-2}$ | -1.86344993528×10$^{-2}$ |
|  | Y | 9.22658604804×10$^{-1}$ | 9.22658604804×10$^{-1}$ | 9.22658604804×10$^{-1}$ |
|  | Z | 3.98334135807×10$^{-1}$ | 3.98334135807×10$^{-1}$ | 3.98334135807×10$^{-1}$ |
| S/C2 Velocity (AU/day) | $V_x$ | -1.72244923440×10$^{-2}$ | -1.72419907995×10$^{-2}$ | -1.72419907995×10$^{-2}$ |
|  | $V_y$ | -1.88198725403×10$^{-4}$ | -1.88384533079×10$^{-4}$ | -1.88384533079×10$^{-4}$ |
|  | $V_z$ | -2.71845314386×10$^{-5}$ | -2.72100311132×10$^{-5}$ | -2.72100311132×10$^{-5}$ |
| S/C3 Position (AU) | X | -1.19599845212×10$^{-2}$ | -1.19599845212×10$^{-2}$ | -1.19599845212×10$^{-2}$ |
|  | Y | 9.22711604030×10$^{-1}$ | 9.22711604030×10$^{-1}$ | 9.22711604030×10$^{-1}$ |
|  | Z | 3.98357113784×10$^{-1}$ | 3.98357113784×10$^{-1}$ | 3.98357113784×10$^{-1}$ |
| S/C3 Velocity (AU/day) | $V_x$ | -1.72249891952×10$^{-2}$ | -1.72424881557×10$^{-2}$ | -1.72424881557×10$^{-2}$ |
|  | $V_y$ | -9.59855278460×10$^{-5}$ | -9.60776184172×10$^{-5}$ | -9.60776184172×10$^{-5}$ |
|  | $V_z$ | -9.55537821052×10$^{-5}$ | -9.56487660660×10$^{-5}$ | -9.56487660660×10$^{-5}$ |

*2.2. The mission orbit optimization*

The goal of the eLISA/NGO mission orbit optimization is to equalize the three arm lengths of the eLISA/NGO formation and to reduce the relative line-of-sight velocities between three pairs of spacecraft as much as possible. In the solar system, the eLISA/NGO spacecraft orbits are perturbed by the planets. With the initial states of the three spacecraft as listed in column three of table 2, we calculated the eLISA/NGO orbit configuration for 1000 days using CGC 2.7. The variations of arm lengths and velocities in the line of sight direction are drawn in figure 2. The largest variations are caused by Earth, Jupiter and Venus. Our method of optimization is to modify the initial velocities and initial heliocentric distances so that (i) the perturbed orbital periods for 1000-day average remains close to one another, and (ii) the average major axes are adjusted to make arms nearly equal. We do this iteratively as follows. From figure 2, we notice that the variation of Arm1 (between S/C2 and S/C3) is small. First, we adjust the initial conditions of S/C2 and S/C3 to make the variation of Arm1 satisfy the mission requirements that arm length variations are within 2 % and Doppler velocities are within 10 m/s. Then we adjust the initial conditions of S/C1 so that Arm2 and Arm3 satisfy the mission requirements. Adjustments are always performed in the ecliptic heliocentric coordinate system.

The actual adjustment procedure is described as follows. Firstly, the magnitudes of initial velocities of S/C2 and S/C3 were adjusted so that their average periods (367.474 days) in 3 years were a little bit longer than 1 sidereal year. Within a definite range, when the periods become longer, the variations of Arm1 become smaller. The initial velocities were adjusted so that the Arm1 satisfied the eLISA/NGO arm length and Doppler velocity requirements. After this, we adjusted the initial velocities of S/C1 to make its orbital period approach those of S/C2 and S/C3, and Arm2 and Arm3 nearly equal. If the results obtained from the above procedure did not satisfy the requirements or better results were expected, we could adjust the orbital periods of S/C2 and S/C3 a little bit longer again under the constraint that the eLISA/NGO requirements for Arm1 is satisfied. Up to this stage, only initial velocities have been adjusted. After we have completed this stage, the initial conditions of the 3 S/C are listed in column 4 of table 2; the variations of arm lengths and velocities in the line of sight direction are drawn in figure 3.

After the first stage, we optimized the orbital period of S/C1 by adjusting the initial velocity and the semi-major axis until the eLISA/NGO requirements were satisfied. The initial conditions of the 3 S/C, after optimization, are listed in column 5 of table 2; the variations of arm lengths (within 2 %) and velocities in the line of sight direction (within 5.5 m/s, better than the less than 10 m/s requirement) are drawn in figure 4. In figure 4, we also draw the angle between barycentre of S/C and Earth in 1000 days; it starts at 10° behind Earth and varies between 9° and 16° with a quasi-period of variation about 1 sidereal year mainly due to Earth's elliptic motion.

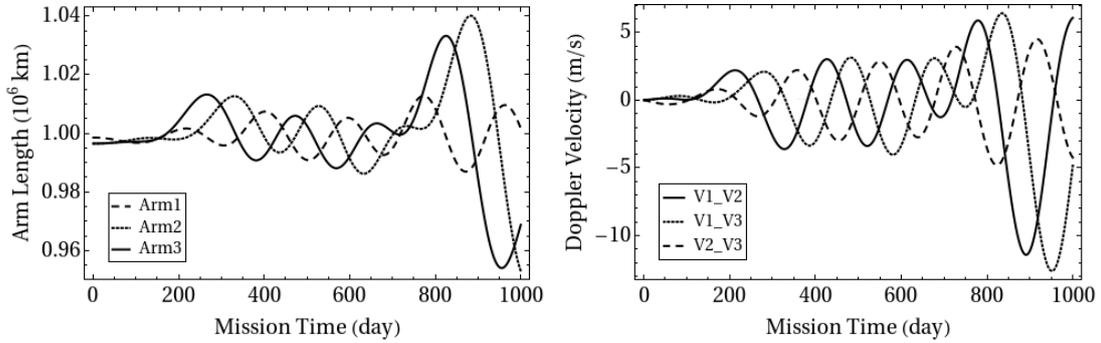

**Figure 2.** Variations of the arm lengths and the velocities in the line of sight direction in 1000 days for the S/C configuration with initial conditions given in column 3 (initial choice) of table 2.

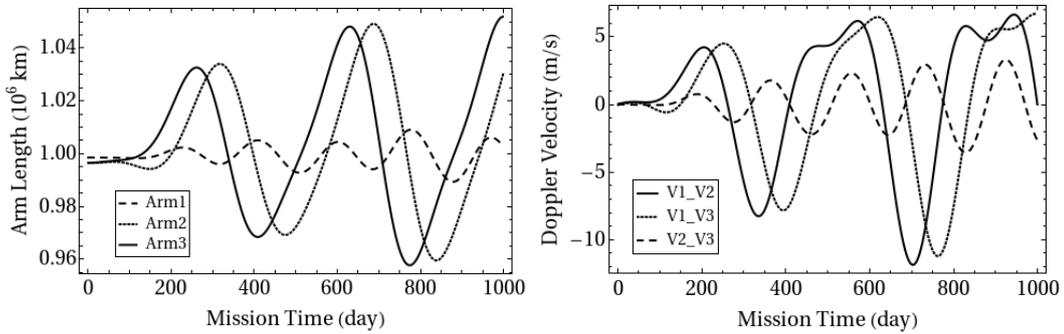

**Figure 3.** Variations of the arm lengths and the velocities in the line of sight direction in 1000 days for the S/C configuration with initial conditions given in column 4 (after the first stage optimization) of table 2.

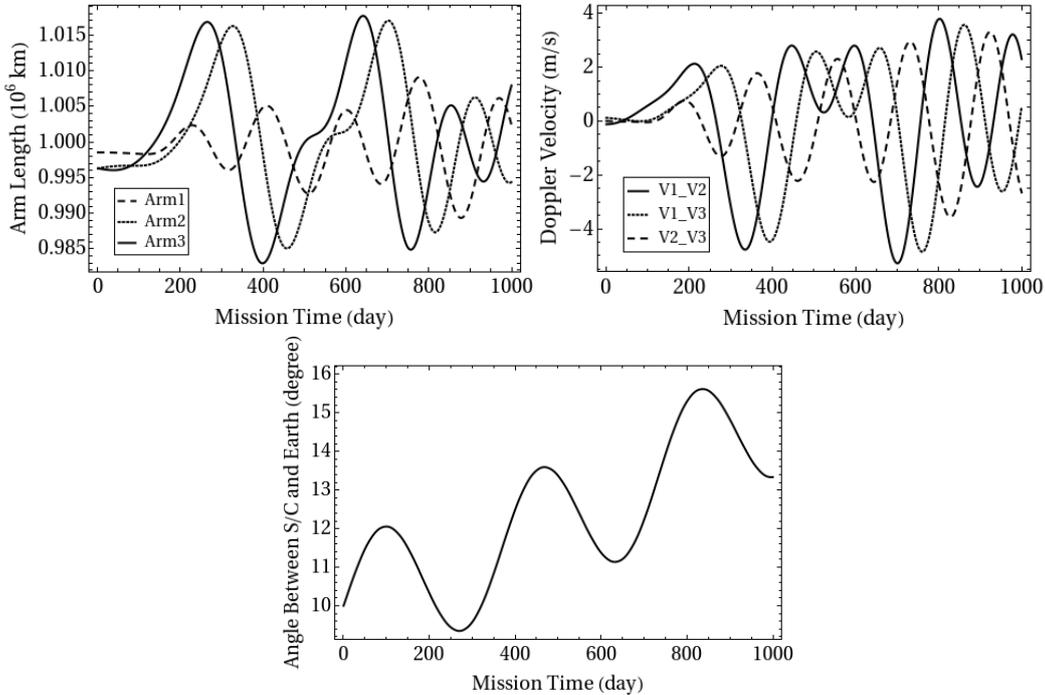

**Figure 4.** Variations of the arm lengths, the velocities in the line of sight direction, and the angle between barycentre of S/C and Earth in 1000 days for the S/C configuration with initial conditions given in column 5 (after final optimization) of table 2.

## 3. Orbit design of an NGO-LISA-type mission with $2 \times 10^6$ km arm length

In the initial choice for initial conditions, we set e = 0.003850; ε = 0.00666 and followed the equations (2.1)-(2.6) in section 2.1 to obtain the initial conditions listed in column 3 of table 3. In the first-stage optimization, we followed the same procedure as in section 2 and the results are listed in column 4 of table 3. However, by adjusting the period and semi-major axis of S/C1, we were not able to find suitable results for Arm2 and Arm3 which satisfied the mission requirement. So we adjusted the initial velocities of S/C2 and S/C3 to extend their period with the constraint that Arm1 satisfying the requirement. After this, we followed the same procedure as in section 2 to adjust both the period and semi-major axis of S/C1 to obtain initial conditions listed in column 5 of table 3 which satisfy the requirements. Variations of the arm lengths and the velocities in the line of sight direction in 1000 days for the S/C configurations with their choices of initial conditions are drawn in figure 5, 6 and 7 respectively. One could notice some similarity between figure 4 and 7; the absolute scale is more or less doubled. In figure 7, we also draw the angle between barycentre of S/C and Earth in 1000 days; as in the case for eLISA/NGO, it starts at 10° behind Earth and varies between 9° and 16° with a quasi-period of variation about 1 sidereal year mainly due to Earth's elliptic motion.

**Table 3.** Initial states (conditions) of 3 S/C of an NGO-LISA-type mission with $2 \times 10^6$ km arm length at epoch JD2459215.5 for our initial choice, after period optimization, and after all optimizations in J2000 equatorial (Earth mean equator and equinox coordinates) solar-system-barycentric coordinate system

|  |  | Initial choice of S/C initial states | Initial states of S/C after 1st stage optimization | Initial states of S/C after final optimization |
|---|---|---|---|---|
| S/C1 Position (AU) | X | $-1.53387344715 \times 10^{-2}$ | $-1.53387344715 \times 10^{-2}$ | $-1.53386649716 \times 10^{-2}$ |
|  | Y | $9.23766504482 \times 10^{-1}$ | $9.23766504482 \times 10^{-1}$ | $9.23759158093 \times 10^{-1}$ |
|  | Z | $4.07903871486 \times 10^{-1}$ | $4.07903871486 \times 10^{-1}$ | $4.07900628141 \times 10^{-1}$ |
| S/C1 Velocity (AU/day) | $V_x$ | $-1.71422212505 \times 10^{-2}$ | $-1.71420961485 \times 10^{-2}$ | $-1.71600223710 \times 10^{-2}$ |
|  | $V_y$ | $-1.41436877401 \times 10^{-4}$ | $-1.41435884024 \times 10^{-4}$ | $-1.41578227841 \times 10^{-4}$ |
|  | $V_z$ | $-6.10850714654 \times 10^{-5}$ | $-6.10846407838 \times 10^{-5}$ | $-6.11463543817 \times 10^{-5}$ |
| S/C2 Position (AU) | X | $-2.19696138442 \times 10^{-2}$ | $-2.19696138442 \times 10^{-2}$ | $-2.19696138442 \times 10^{-2}$ |
|  | Y | $9.22386466148 \times 10^{-1}$ | $9.22386466148 \times 10^{-1}$ | $9.22386466148 \times 10^{-1}$ |
|  | Z | $3.96438156600 \times 10^{-1}$ | $3.96438156600 \times 10^{-1}$ | $3.96438156600 \times 10^{-1}$ |
| S/C2 Velocity (AU/day) | $V_x$ | $-1.72403916985 \times 10^{-2}$ | $-1.72403916985 \times 10^{-2}$ | $-1.72579801919 \times 10^{-2}$ |
|  | $V_y$ | $-2.34559090306 \times 10^{-4}$ | $-2.34559090306 \times 10^{-4}$ | $-2.34792996947 \times 10^{-4}$ |
|  | $V_z$ | $7.09717982718 \times 10^{-6}$ | $7.09717982718 \times 10^{-6}$ | $7.10656036726 \times 10^{-6}$ |
| S/C3 Position (AU) | X | $-8.60799470447 \times 10^{-3}$ | $-8.60799470447 \times 10^{-3}$ | $-8.60799470447 \times 10^{-3}$ |
|  | Y | $9.22492564568 \times 10^{-1}$ | $9.22492564568 \times 10^{-1}$ | $9.22492564568 \times 10^{-1}$ |
|  | Z | $3.96484155895 \times 10^{-1}$ | $3.96484155895 \times 10^{-1}$ | $3.96484155895 \times 10^{-1}$ |
| S/C3 Velocity (AU/day) | $V_x$ | $-1.72413863435 \times 10^{-2}$ | $-1.72413706152 \times 10^{-2}$ | $-1.72589601077 \times 10^{-2}$ |
|  | $V_y$ | $-4.98816144912 \times 10^{-5}$ | $-4.98815738724 \times 10^{-5}$ | $-4.99269993859 \times 10^{-5}$ |
|  | $V_z$ | $-1.29946678724 \times 10^{-4}$ | $-1.29946562046 \times 10^{-4}$ | $-1.30077047795 \times 10^{-4}$ |

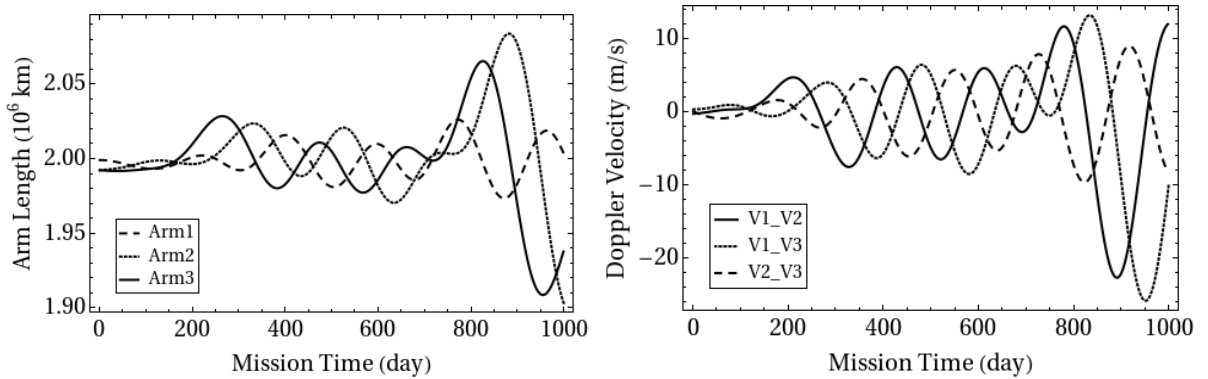

**Figure 5.** Variations of the arm lengths and the velocities in the line of sight direction in 1000 days for the S/C configuration with initial conditions given in column 3 (initial choice) of table 3.

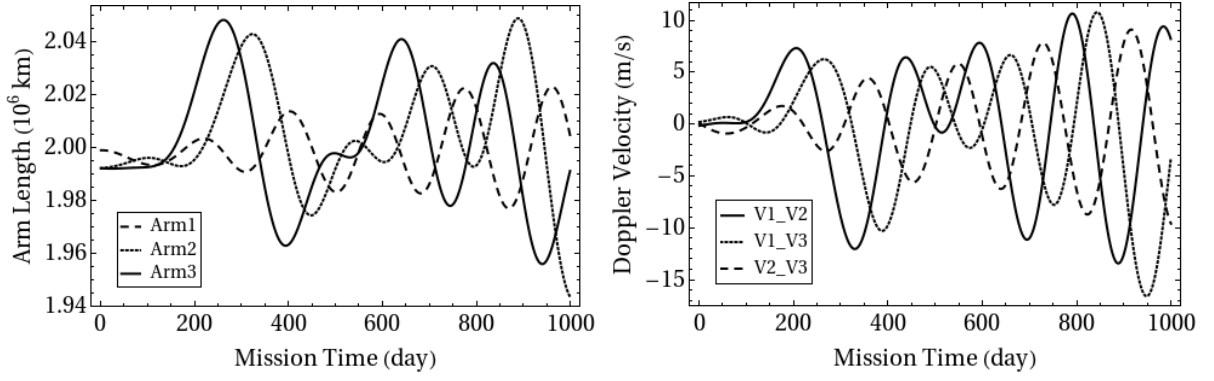

**Figure 6.** Variations of the arm lengths and the velocities in the line of sight direction in 1000 days for the S/C configuration with initial conditions given in column 4 (after first stage optimization) of table 3.

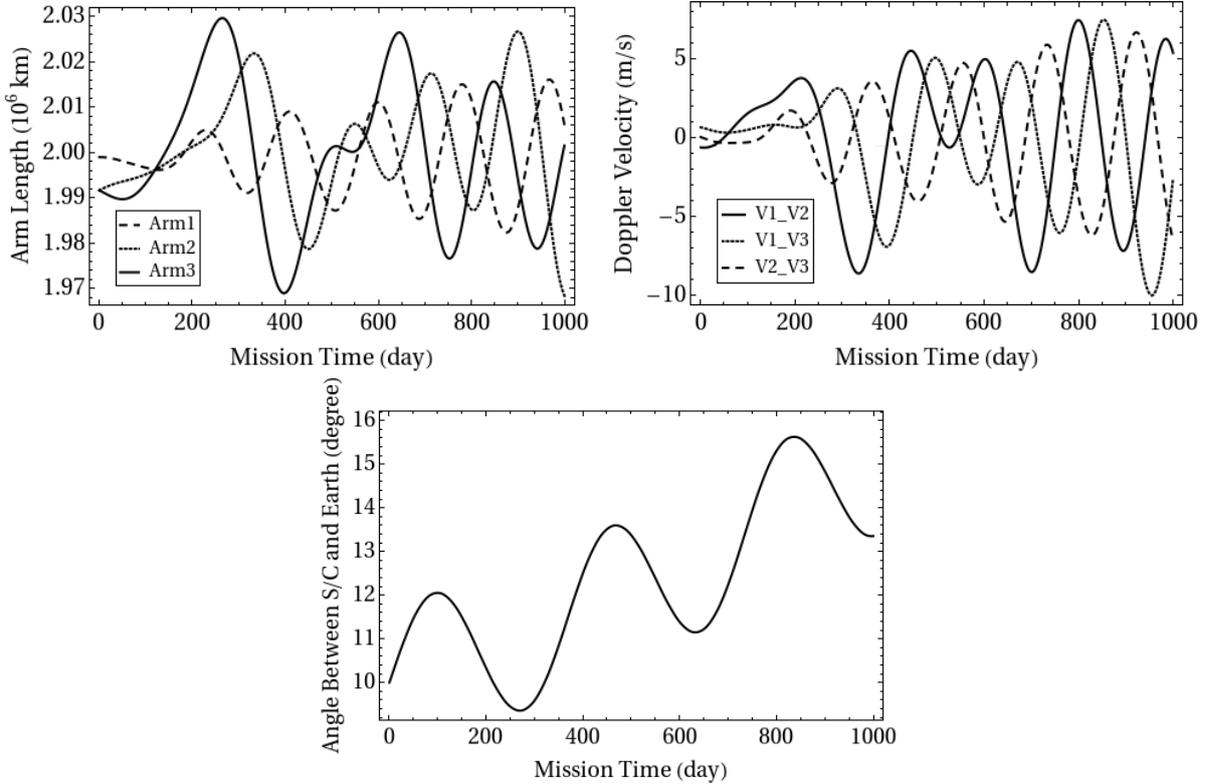

**Figure 7.** Variations of the arm lengths, the velocities in the line of sight direction and angle between barycentre of S/C and Earth in 1000 days for the S/C configuration with initial conditions given in column 5 (after optimization) of table 3.

## 4. Numerical simulation of the second-generation TDI for eLISA/NGO

In our previous papers (Dhurandhar, Ni and Wang, 2013; Wang and Ni, 2013), we have used the CGC 2.7 ephemeris framework to calculate the difference between the two path lengths for TDI configurations obtained by Dhurandhar *et al* (2010). The results were showed by plotting the difference as function of the epoch of LISA and ASTROD-GW orbit configuration respectively. These TDI configurations belong to a large family of the second-generation analytic solutions of time delay interferometry with one arm dysfunctional. The method of obtaining these solutions and the TDI configurations were briefly reviewed in section 1.

In the numerical calculation, we used the CGC 2.7 ephemeris framework to calculate the difference between the two path lengths for TDI configurations and plotted the difference as function of the signal arriving epoch of TDI in the eLISA/NGO's orbit, i.e., the difference of starting time was calculated. We made use of the iteration and interpolation methods (Chiou and Ni, 2000a, 2000b; Newhall, 1989; Li and Tian, 2004) to calculate the time in the barycentric coordinate system. The results for n = 1, 2, 3 configurations of eLISA/NGO are shown in figure 8, 9, 10 respectively and tabulated in table 1. From the last diagram in figure 9, we noticed that, the accuracy of this calculation should be better than 1 μm (3.3 fs) for the path difference (whether we include relativistic light propagation or not). We also noted that all the time differences are below 43 ps (corresponding to the maximum path length difference of 13 mm). This is well below the limit for TDIs which the laser frequency noise is required to be suppressed.

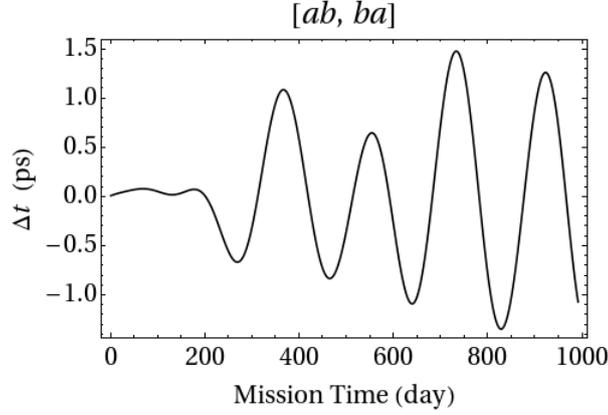

**Figure 8.** The difference of the two optical path lengths for n = 1 TDI configuration.

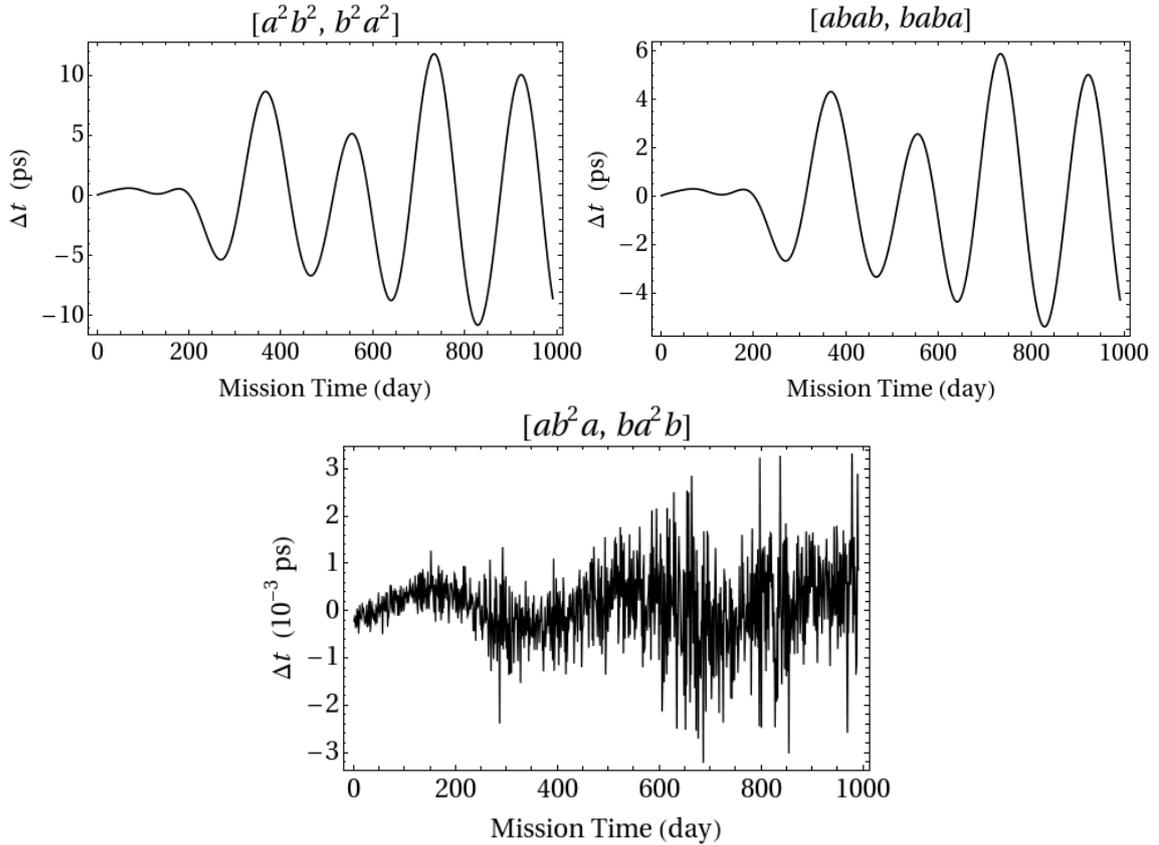

**Figure 9.** The difference of the two optical paths lengths for three n = 2 TDI configurations.

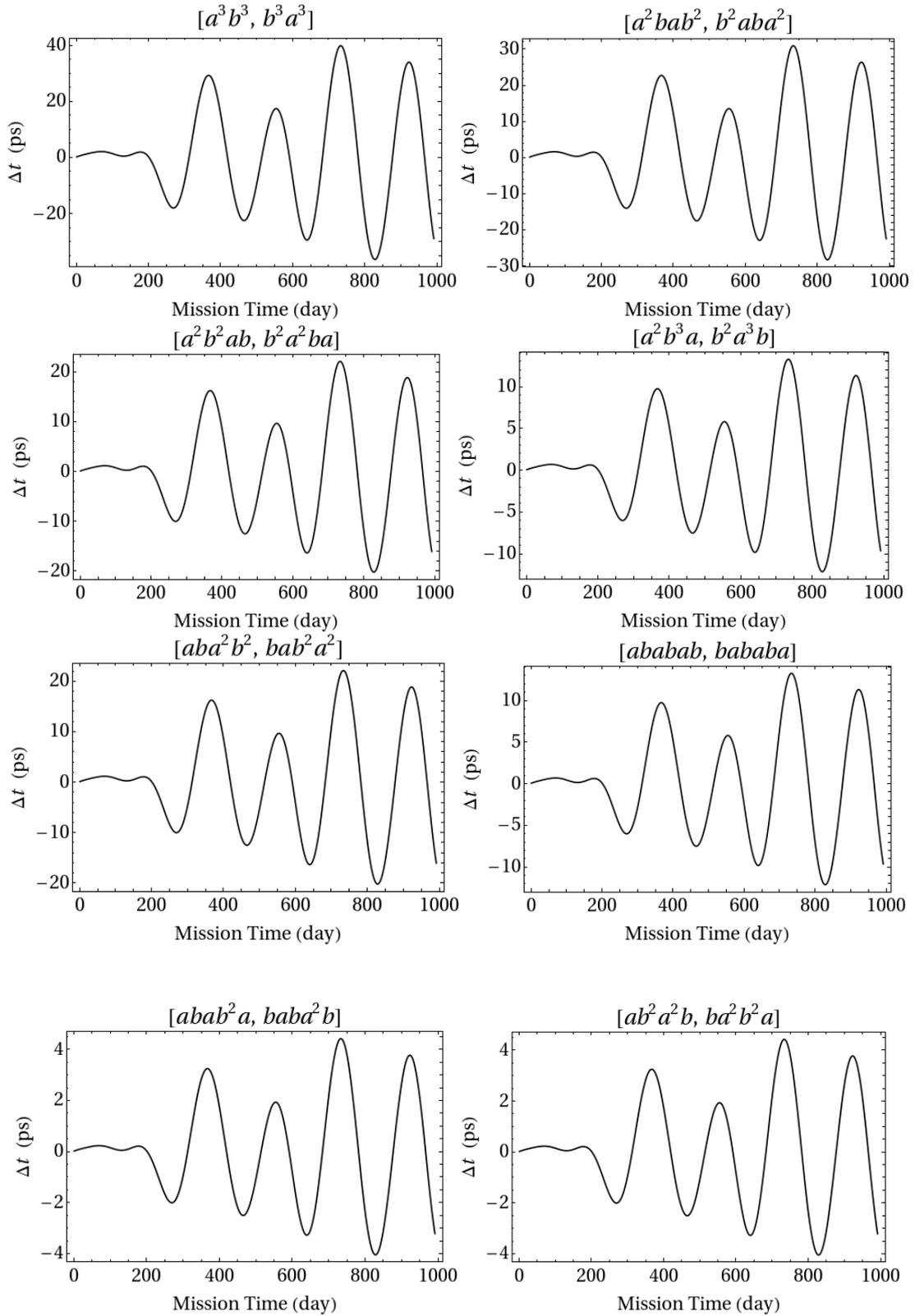

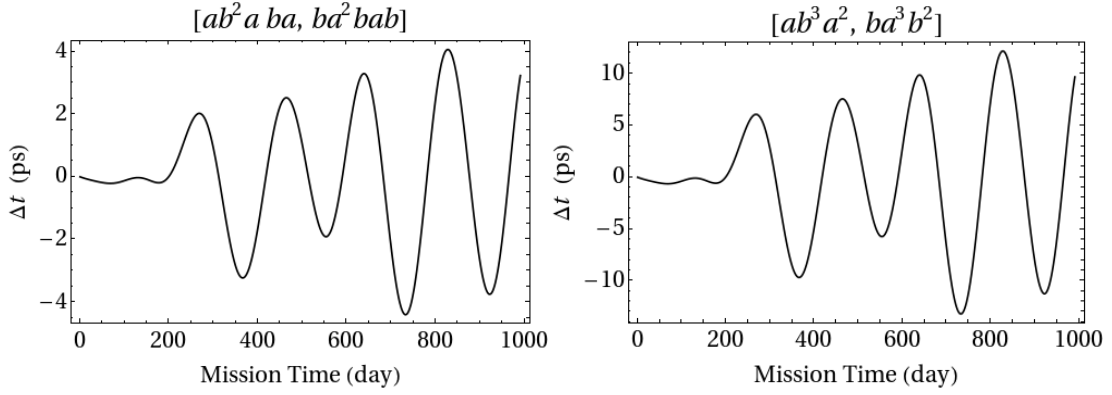

**Figure 10.** The difference of the two optical path lengths for ten n = 3 TDI configurations.

It is interesting to note that for one of the commutators, namely [$ab^2a$, $ba^2b$], the maximal path difference is about $3.3 \times 10^{-15}$ sec, much lower than others. This is the result of cancellation of higher order terms in the time derivatives of $L(t)$. Specifically, there is greater symmetry in this combination in which the $\dddot{L}$ terms also cancel out. As we mentioned in the last paragraph, this figure indicates that the numerical accuracy is better than 3.3 fs (1 μm).

## 5. Numerical simulation of second-generation TDI for an NGO-LISA-type mission with $2 \times 10^6$ km arm length

In this section, we do the same thing as in the last section for an NGO-LISA-type mission with $2 \times 10^6$ km arm length. The results are shown in figure 11-13 and tabulated in table 1.

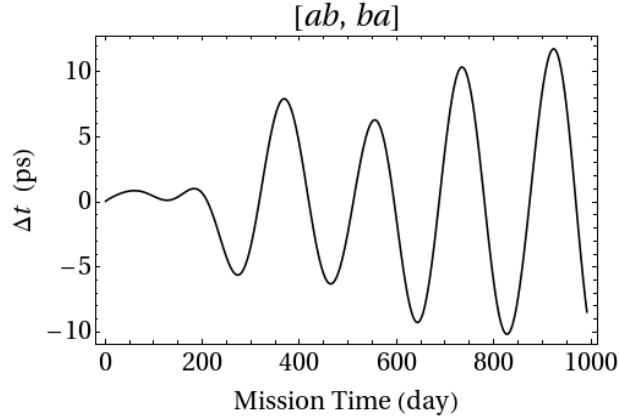

**Figure 11.** The difference of two optical path lengths for n = 1 TDI configuration.

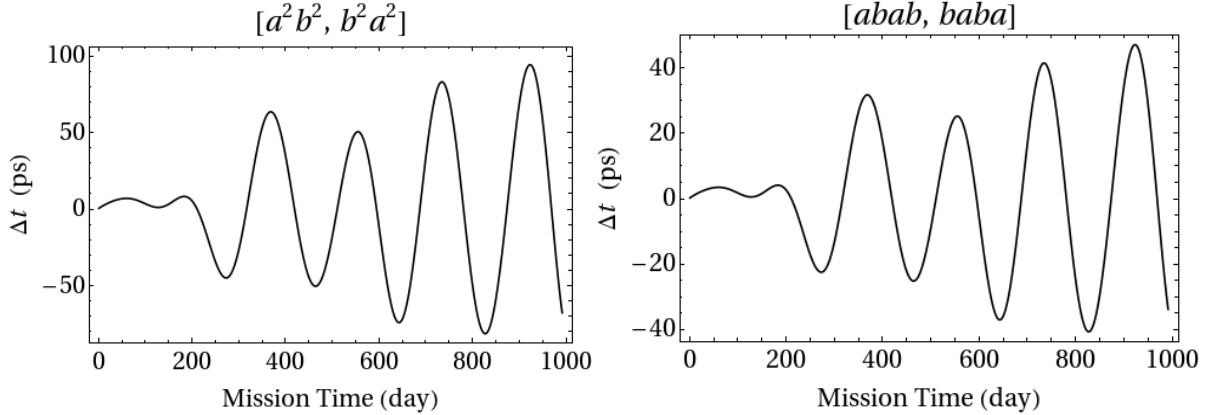

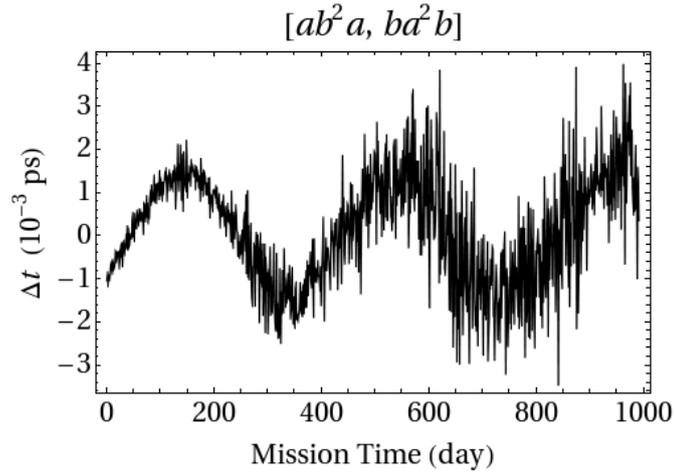

**Figure 12.** The difference of two optical path lengths for three n = 2 TDI configurations.

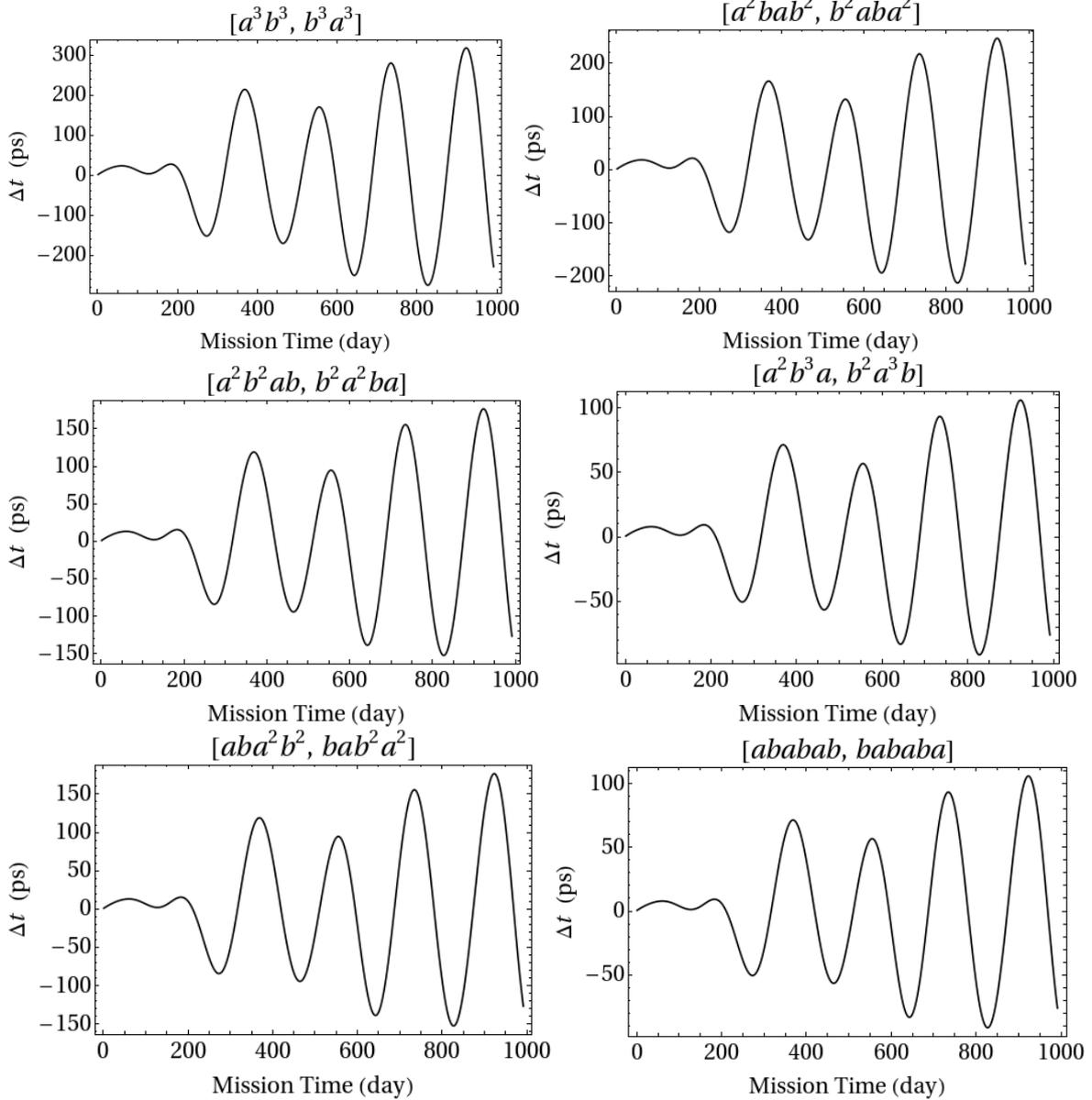

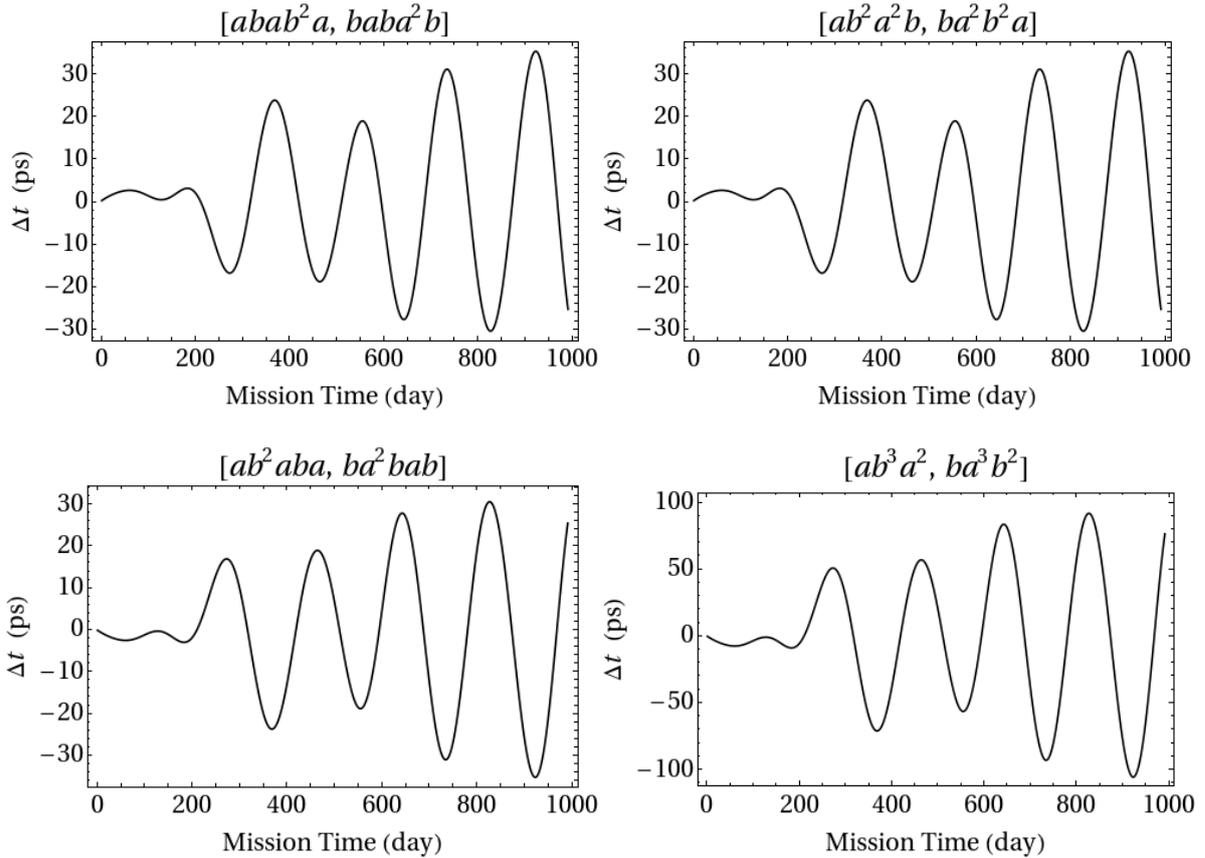

**Figure 13.** The difference of two optical path lengths for ten n = 3 TDI configurations.

## 6. Discussion and outlook

In the eLISA/NGO mission for detecting GWs in the frequency range 100 μHz – 1 Hz (30 μHz – 1 Hz *as a goal*), the 2 "daughter" S/C range interferometrically with the "mother" S/C with arm lengths of about 1 million kilometers. After optimization of the mission orbit, the changes of arm length can be less than 20,000 km and the relative Doppler velocities can be less than ±5.5 m/s for 1000 days. In order to attain the requisite sensitivity for NGO, laser frequency noise must be suppressed to below the secondary noises such as the optical path noise, acceleration noise etc. For suppressing the laser frequency noise, the second-generation TDI configurations satisfy the requirements in the general case of flexing arms. We worked out a set of 1000-day optimized mission orbits of NGO spacecraft and calculated the residual errors in the second-generation TDI in this paper. We have examined a total of 14 such TDI solutions which may be deemed sufficient for the purpose of astrophysical observations. All the second-generation TDIs calculated in this paper have optical path differences below 13 mm and well satisfy the NGO time offset requirement of 10 m (i.e., about 30 ns). The numerical method used here could be readily applied to other 2-arm interferometers, and we have applied to an NGO-LISA-type mission with a nominal arm length of $2 \times 10^6$ km since extension of arm length may be desirable if there are more funding/more countries involved. We have also compared and discussed the resulting differences due to different arm lengths for various mission proposals -- eLISA/NGO, an NGO-LISA-type mission with a nominal arm length of $2 \times 10^6$ km, LISA and ASTROD-GW.

**Acknowledgements**

We would like to thank the National Science Council (Grants No. NSC100-2119-M-007-008 and No. NSC100-2738-M-007-004) and the National Center of Theoretical Sciences (NCTS) for supporting this work in part. We would also like to thank the referees for their constructive comments.

# References

Ando M and the DECIGO Working Group 2013 DECIGO Pathfinder Mission *Int. J. Mod. Phys. D* **22** 1341002 (10 pages)

Arun K G, Iyer B R and Ni W-T 2012 Fifth ASTROD Symposium and Outlook of Direct Gravitational-Wave Detection *Asia Pacific Physics Newsletter (APPN)* **1** (2) 6-11

Austermann J E *et al* 2012 SPTpol: an instrument for CMB polarization measurements with the South Pole Telescope, *Proc. SPIE 8452, Millimeter, Submillimeter, and Far-Infrared Detectors and Instrumentation for Astronomy VI*, 84520E; arXiv:1210.4970

Braxmaier C *et al* 2012 Astrodynamical Space Test of Relativity using Optical Devices I (ASTROD I)—a class-M fundamental physics mission proposal for cosmic vision 2015–2025: 2010 Update *Exp. Astron.*, **34** 181 [arXiv:1104.0060]; and references therein

Brumberg V A 1991 Essential Relativistic *Celestial Mechanics* (Bristol: Adam Hilger) pp 176-177

Chiou D-W and Ni W-T 2000a ASTROD Orbit Simulation and Accuracy of Relativistic Parameter Determination *Advances in Space Research* **25** 1259-1262

Chiou D-W and Ni W-T 2000b Orbit Simulation for the Determination of Relativistic and Solar-System Parameters for the ASTROD Space Mission, *presented to 33rd COSPAR Scientific Assembly, Warsaw, 16-23 July, 2000* [arXiv:astro-ph/0407570]

Crowder J and Cornish N J 2005 Beyond LISA: Exploring Future Gravitational Wave Missions *Phys. Rev. D* **72** 083005; and references therein

Cutler C 1998 Angular resolution of the LISA gravitational wave detector *Phys. Rev. D* **57** 7089-7102

Demorest P *et al* 2009 Gravitational Wave Astronomy Using Pulsars: Massive Black Hole Mergers & the Early Universe *white paper submitted to the Astro2010 Decadal Survey* [arXiv:0902.2968]; and references therein

Dhurandhar S V *et al* 2005 Fundamentals of the LISA stable flight formation *Class. Quantum Grav.* **22** 481-487

Dhurandhar S V *et al* 2010 Time Delay Interferometry for LISA with one arm dysfunctional *Class. Quantum Grav.* **27** 135013

Dhurandhar S V, Ni W-T and Wang G 2013 Numerical simulation of time delay interferometry for a LISA-like mission with the simplification of having only one interferometer, *Adv. Space Res.* **51** 198-206; arXiv:1102.4965

Folkner W M, Hechler F, Sweetser T H, Vincent M A and Bender P L 1997 LISA orbit selection and stability *Class. Quantum Grav.* **14** 1405-1410

George E M *et al* 2012 Performance and on-sky optical characterization of the SPTpol instrument *Proc. SPIE 8452, Millimeter, Submillimeter, and Far-Infrared Detectors and Instrumentation for Astronomy VI*, 84521F [arXiv:1210.4971]

Hechler F and Folkner W M 2003 Mission analysis for the Laser Interferometer Space Antenna (LISA) mission *Adv. Space Res.* **32** 1277-1282

http://eLISA-ngo.org/
http://universe.nasa.gov/new/program/bbo.html
http://www.icrr.u-tokyo.ac.jp/2012/01/28161746.html

Hughes S P 2002 Preliminary optimal orbit design for laser interferometer space antenna *25th Annual AAS Guidance and Control Conference (Breckenridge CO, Feb. 2002)*

Jennrich O *et al* 2011 NGO (New Gravitational wave Observatory) assessment study report, ESA/SRE(2011)19

Kawamura S *et al* 2006 The Japanese space gravitational wave antenna—DECIGO *Class. Quantum Grav.* **23** S125-S131

Kawamura S et al 2011 The Japanese space gravitational wave antenna: DECIGO *Class. Quantum Grav.* **28** 094011 (12pp)

Krolak A *et al* 2004 Optimal filtering of the LISA data *Phys. Rev. D* **70** 022003

Kuroda K (on behalf of the LCGT Collaboration) 2010 Status of LCGT *Class. Quantum Grav.* **27** 084004 (8pp)

Li G and Tian L 2004 PMOE 2003 Planetary Ephemeris Framework (V) Creating and using of Ephemeris Files (in Chinese) *Publications of Purple Mountain Observatory* **23** 160-170

Li G *et al* 2008 Methods for orbit optimization for the LISA gravitational wave observatory *Int. J. Mod.*

 *Phys. D* **17** 1021-1042
LISA Study Team 2000 LISA (Laser Interferometer Space Antenna): A Cornerstone Mission for the Observation of Gravitational Waves; See the sites
 http://sci.esa.int/science-e/www/area/index.cfm?fareaid=27; http://lisa.gsfc.nasa.gov
McMahon J *et al* 2012 Multi-Chroic Feed-Horn Coupled TES Polarimeters *submitted to the proceedings of Low Temperature Detectors-14 (LTD14)*; arXiv:1201.4124
Men J-R *et al* 2009 ASTROD-GW mission orbit design *Proceedings of Sixth Deep Space Exploration Technology Symposium, December 3-6, 2009, Sanya, Hainan, China* 47-52
Men J-R *et al* 2010 Design of ASTROD-GW orbit *Chinese Astronomy and Astrophysics* **34** 434-446
Newhall X X 1989 Numerical Representation of Planetary Ephemerides *Celestial Mechanics* **45** 305-310
Ni W-T *et al* 1997 Progress in Mission Concept Study and Laboratory Development for the ASTROD Astrodynamical Space Test of Relativity using Optical Devices, *Proceedings of SPIE 3116: Small Spacecraft, Space Environments, and Instrumentation Technologies, ed. F. A. Allahdadi, E. K. Casani, and T. D. Maclay*, pp.105-116, SPIE
Ni W-T 2008 ASTROD and ASTROD I -- Overview and Progress *Int. J. Mod. Phys. D* **17** 921-940; and references therein
Ni W-T 2009a Super-ASTROD: Probing primordial gravitational waves and mapping the outer solar system *Class. Quantum Grav.* **26** 075021; and references therein
Ni W-T 2009b ASTROD Optimized for Gravitational-wave Detection: ASTROD-GW – *a pre-Phase A study proposal submitted to Chinese Academy of Sciences February 26*
Ni W-T 2010 Gravitational waves, dark energy and inflation *Mod. Phys. Lett. A* **25** 922-935 [arXiv:1003.3899]
Ni W-T 2012 Dark energy, co-evolution of massive black holes with galaxies, and ASTROD-GW, Paper (COSPAR paper number H05-0017-10) presented in the 38th COSPAR Scientific Assembly, 18-25 July 2010, Bremen, Germany, *Adv. Space Res.*, http://dx.doi.org/10.1016/j.asr.2012.09.019; arXiv:1104.5049
Ni W-T 2013 ASTROD-GW: Overview and Progress *Int. J. Mod. Phys. D* **22** 1341004
Ni W-T *et al* 2009 ASTROD optimized for gravitational wave detection: ASTROD-GW *Proceedings of Sixth Deep Space Exploration Technology Symposium, December 3-6, 2009, Sanya, Hainan, China* 122-128
Niemack M D *et al* 2010 ACTPol: A polarization-sensitive receiver for the Atacama Cosmology Telescope, *Proc. SPIE* **7741** 77411S; arXiv:1006.5049
Phinney S *et al* 2004 *The Big Bang Observer: Direct detection of gravitational waves from the birth of the Universe to the Present*, NASA Mission Concept Study
Planck Surveyor 2012, www.rssd.esa.int/index.php?project=PLANCK
Soffel M *et al* 2003 The IAU 2000 resolutions for astrometry, celestial mechanics and metrology in the relativistic framework: explanatory supplement *Astron J* **126** 2687-2706
Tang C-J and Ni W-T 2000 Asteroid Perturbations and Mass Determination for the ASTROD Space Mission, *presented to the 33$^{rd}$ COSPAR Scientific Assembly, Warsaw, 16-23 July, 2000*; arXiv:astro-ph/0407606
Tang C-J and Ni W-T 2002 CGC 2 ephemeris framework *Publications of the Yunnan Observatory* **3** 21-32
The Advanced LIGO Team 2012, http://www.ligo.caltech.edu/advLIGO/
The Advanced Virgo Team 2012, http://wwwcascina.virgo.infn.it/advirgo/
Tinto M and Dhurandhar S V 2005 Time-Delay Interferometry *Living Rev. Relativity* **8** 4; and references therein
Vincent M A and Bender P L 1987 *Proc. Astrodynamics Specialist Conference* (Kalispell USA) (Univelt, San Diego) p 1346
Wang G 2011 Time-delay interferometry for ASTROD-GW *Master thesis* Purple Mountain Observatory
Wang G and Ni W-T 2011 ASTROD-GW time delay interferometry (in Chinese) *Acta Astron. Sin.* **52** 427-442
Wang G and Ni W-T 2012 ASTROD-GW time delay interferometry *Chin. Astron. Astrophys.* **36** 211-228
Wang G and Ni W-T 2013 Orbit optimization for ASTROD-GW and its time delay interferometry with two arms using CGC ephemeris *Chinese Physics B, in press*, arXiv:1205.5175
Yi Z *et al* 2008 The design of the orbits of the LISA spacecraft *Int. J. Mod. Phys. D* **17** 1005-1019